\documentclass[twocolumn]{aastex62}
\usepackage{xspace}
\definecolor{DarkGreen}{rgb}{0.0,0.4,0.0}  

\newcommand{\kms}{km\,s$^{-1}$\xspace}
\shorttitle{Filament Splitting}
\shortauthors{Pan et al.}

\begin{document}

\title{Pre-eruption Splitting of the Double-Decker Structure in a Solar Filament}

\correspondingauthor{Rui Liu}
\email{rliu@ustc.edu.cn}

\author[0000-0002-0318-8251]{Hanya Pan}
\affiliation{CAS Key Laboratory of Geospace Environment, Department of Geophysics and Planetary Sciences, University of Science and Technology of China, Hefei 230026, China}

\author[0000-0003-4618-4979]{Rui Liu}
\affiliation{CAS Key Laboratory of Geospace Environment, Department of Geophysics and Planetary Sciences, University of Science and Technology of China, Hefei 230026, China}
\affiliation{CAS Center for Excellence in Comparative Planetology, Hefei 230026, China}
\affiliation{Mengcheng National Geophysical Observatory, School of Earth and Space Sciences, University of Science and Technology of China, Hefei 230026, China}

\author[0000-0003-0510-3175]{Tingyu Gou}
\affiliation{CAS Key Laboratory of Geospace Environment, Department of Geophysics and Planetary Sciences, University of Science and Technology of China, Hefei 230026, China}
\affiliation{CAS Center for Excellence in Comparative Planetology, Hefei 230026, China}

\author[0000-0002-5740-8803]{Bernhard Kliem}
\affiliation{Institute of Physics and Astronomy, University of Potsdam, 14476 Potsdam,
Germany}

\author[0000-0001-9647-2149]{Yingna Su}
\affiliation{Key Laboratory of DMSA, Purple Mountain Observatory, Chinese Academy of Sciences, Nanjing 210008, China}
\affiliation{School of Astronomy and Space Science, University of Science and Technology of China, Hefei, 230026, China}

\author[0000-0003-3060-0480]{Jun Chen}
\affiliation{CAS Key Laboratory of Geospace Environment, Department of Geophysics and Planetary Sciences, University of Science and Technology of China, Hefei 230026, China}
\affiliation{Institute of Physics and Astronomy, University of Potsdam, 14476 Potsdam, Germany}

\author[0000-0002-8887-3919]{Yuming Wang}
\affiliation{CAS Key Laboratory of Geospace Environment, Department of Geophysics and Planetary Sciences, University of Science and Technology of China, Hefei 230026, China}
\affiliation{CAS Center for Excellence in Comparative Planetology, Hefei 230026, China}


\begin{abstract}
Solar filaments often erupt partially. Although how they split remains elusive, the splitting process has the potential of revealing the filament structure and eruption mechanism. Here we investigate the pre-eruption splitting of an apparently single filament and its subsequent partial eruption on 2012 September 27. The evolution is characterized by three stages with distinct dynamics. During the quasi-static stage, the splitting proceeds gradually for about 1.5 hrs, with the upper branch rising at a few kilometers per second and displaying swirling motions about its axis. During the precursor stage that lasts for about 10 min, the upper branch rises at tens of kilometers per second, with a pair of conjugated dimming regions starting to develop at its footpoints; with the swirling motions turning chaotic, the axis of the upper branch whips southward, which drives an arc-shaped EUV front propagating in the similar direction. During the eruption stage, the upper branch erupts with the onset of a C3.7-class two-ribbon flare, while the lower branch remains stable. Judging from the well separated footpoints of the upper branch from those of the lower one, we suggest that the pre-eruption filament processes a double-decker structure composed of two distinct flux bundles, whose formation is associated with gradual magnetic flux cancellations and converging photospheric flows around the polarity inversion line. 

\end{abstract}



\section{Introduction} \label{sec:intro}
Solar filaments (or \emph{prominences} if observed above the limb) are cold and dense plasma `clouds' suspended in the corona. They exhibit a broad spectrum of eruptive dynamics \citep[][and references therein]{Gilbert2007}, the two extremes of which are: 1) \emph{full eruptions}, in which filament mass and associated magnetic structure completely escapes the Sun, and 2) \emph{failed eruptions}, in which the filament is first activated as if a successful eruption is ongoing but the eruptive process is suddenly halted in the low corona, with none of the filament mass nor magnetic structure escaping the Sun. The majority between the two extremes are \emph{partial eruptions}---only part of the magnetic structure and/or filament mass is expelled. Often the filament splits dynamically during the eruption \citep[e.g.,][]{Liu2007,Tripathi2009,Cheng2018}, exhibiting signatures of internal magnetic reconnection \citep{Gibson&Fan2006}. Filaments may also split before the eruption \citep{Contarino2003,Guo2010,Tian2018}, but how it works is poorly known, due to subtle activity and emission during this phase. However, the splitting may provide a rare opportunity for us to explore the relevant magnetic structure and eruption mechanism.

It has been under debate as to whether the dense filament mass is supported against gravity by magnetic tension force in dipped field lines at the bottom of a flux rope or on the top of a sheared arcade \citep{Mackay2010,Liu2020}. Since a flux-rope configuration is often detected in situ when interplanetary CMEs pass through spacecrafts \citep{Burlaga1981}, pre-eruptive structures like filaments \citep[especially the prominence-cavity system;][]{Gibson2015} and sigmoids \citep[S-shaped emission in soft X-ray or EUV;][]{Rust&Kumar1996,Liu2010tc,Zhang2012} are proposed to be indicators of flux ropes on the Sun. Naturally, ideal magnetohydrodynamic instabilities of current-carrying flux ropes, especially the torus instability \citep[e.g.,][]{Kliem&Torok2006,Torok&Kliem2007,Fan2010,WangD2017} and the helical kink instability \citep[e.g.,]{Fan2005,Torok&Kliem2005}, have been intensively investigated as candidate eruption mechanisms. The torus instability results from an imbalance between two competing forces with increasing heights, where the upward Lorentz self-force of the flux-rope current (also known as the `hoop' force) dominates over the downward tension force of the external `strapping' field. The helical kink instability converts magnetic twist into writhe to relax the tension in the twisted field, which is often manifested as a filament rotates into an inverted $\gamma$ or $\delta$ shape in projection \citep[e.g.,][]{Gilbert2007}. A kink-unstable flux-rope can be confined in the low corona if it is in the torus-stable regime \citep{Torok&Kliem2005}.

Since we still cannot measure the coronal magnetic field, we may take filament mass as a field tracer, especially when high-resolution data is available to resolve the individual threads and to monitor their dynamic motions \citep[e.g.,][]{Awasthi2019}. In particular, investigations into ``double-decker'' filaments have shed new light on the magnetic structure of filaments. Two possible configurations, a double flux rope or a single flux rope atop a sheared arcade, were proposed by \citet{Liu2012} to explain the double-decker filament consisting of two main branches with clear vertical separation prior to eruption. The dots were soon connected between double-decker filaments and a long-standing puzzle --- a stable filament is often left behind in sigmoid eruptions associated with CMEs \citep[e.g.,][]{Pevtsov2002,LiuC2007,Liu2008,Cheng2014}. This can be explained by a double-decker configuration whose upper branch is void of filament material \citep{Cheng2014}. The non-linear-force-free-field (NLFFF) modeling of coronal magnetic field has demonstrated the possibility of double or even multiple flux bundles piled up along the same polarity inversion line \cite[PIL; e.g.,][]{Regnier&Amari2004, Liu2016, Liu2017, Hou2018, Awasthi2018}. In addition, \citet{Awasthi2019} inferred a double-decker structure with opposite signs of helicity from the dynamic motions in a disturbed filament, i.e., rotation about the spine and longitudinal oscillation along the spine. \citet{ChenJL2020} studied a filament that splits into a C-shaped and a reverse C-shaped branch in a failed eruption. Using the flux-rope-insertion method, they found that a double flux rope fits the observation well: the lower rope has a weak positive twist but the upper rope possesses a strong negative twist. This setting may be more stable than that originally envisaged by \citet{Liu2012}, as the two branches are not separated by a hyperbolic flux tube \citep[HFT; see also][]{ Jelinek2020}, where a current layer likely develops \citep{Titov2002}; but at the same time, it makes relevant the tilt instability between opposing-directed current channels \citep{Keppens2019}. 

A wide range of eruptive behaviors have been reported in the studies of double-decker filaments. Often the upper branch erupts as a coronal mass ejection (CME) with the lower branch being left behind \citep[e.g.,][]{Liu2012, Zhu&Alexander2014,Cheng2014}. Additionally, the two branches may coalesce into a single structure before eruption \citep[e.g.,][]{Zhu2015,Tian2018} or into a single CME after successive eruptions \citep[e.g.,][]{Dhakal2018}. Sometimes filament threads within the lower branch are intermittently brightened and transported into the upper branch during its slow-rise phase \citep{Liu2012,Zhu&Alexander2014}. Through such incremental episodes of flux transfer, the upper branch may become unstable when it acquires excessive axial flux relative to the ambient \citep[e.g.,][]{Su2011,Zhang2014,Zhang2020}. Motivated by observations, \citet{Kliem2014} employed two concentric, toroidal flux ropes to model double-decker filaments. Trough a transfer of flux and current from the lower to the upper rope, they were able to reproduce the eruption of the upper rope with the lower rope being stable. Alternatively, \citet{Kliem2014} identified in an MHD simulation a double flux rope that are stacked closely together until their splitting shortly before the eruption. At the HFT separating the two ropes, tether-cutting reconnection with the ambient field adds flux and twist to the upper rope but strengthens the flux overlying the lower one, which leads to the splitting and partial eruption.

Despite the above observations and simulations, it remains unclear how a double-decker filament forms. \citet{Kliem2014} suggests that the splitting may occur through tether-cutting reconnection shortly before eruption. Relevant observations are extremely rare. Previous studies showed that a single filament may split before eruption \citep{Contarino2003,Guo2010,Tian2018}, but did not report any reconnection signature at the splitting place. Here we study the splitting and subsequent eruption of an intermediate filament between two active regions. With multi-wavelength and stereoscopic observations, this study provides new insights into the formation and eruption mechanism of double-decker filaments. In the sections that follow, we present the detailed analysis of the observations in \S\ref{sec:obs}, and then discuss the interpretations and implications of the observations in \S\ref{sec:disc}.

\section{Observation \& Analysis} \label{sec:obs}

\subsection{Instruments}\label{subsec:ovv}
The filament eruption under investigation is associated with a GOES-class C3.7 flare and a fast halo CME on 2012 September 27. According to GOES 1--8~{\AA} flux, the flare onsets at 23:36 UT and peaks at 23:57 UT. We investigate the eruption mainly using extreme ultraviolet (EUV) images taken by the Atmospheric Imaging Assembly \citep[AIA;][]{Lemen2012} onboard the Solar Dynamics Observatory \citep[SDO;][]{Pesnell2012} and by the Extreme-UltraViolet Imager \citep[EUVI;][]{Wuelser2004} onboard the `Ahead' satellite (hereafter STA) of the Solar Terrestrial Relations Observatory \citep[STEREO;][]{Kaiser2008}. As of the flare peak, the separation angle between STA and Earth is 125.6 deg. Despite the cloudy weather, the filament splitting was observed in the H$\alpha$ line center and blue wing by the Big Bear Solar Observatory (BBSO). 

AIA takes full-disk images with a spatial scale of $0''.6$ pixel$^{-1}$ and a cadence of 12 s. Among its 7 EUV and 2 UV passbands, we fixated our attention on 131~{\AA} (primarily \ion{Fe}{21} for flare plasma, with a peak response temperature $\log T = 7.05$; \ion{Fe}{8} for active regions, $\log T = 5.6$),  193~{\AA} (\ion{Fe}{24} for flare plasma, $\log T = 7.25$; \ion{Fe}{12} for active regions, $\log T = 6.2$), and 304~{\AA} (\ion{He}{2}, $\log T = 4.7$). EUVI's 195 and 304~{\AA} passbands are similar to AIA's 193 and 304~{\AA}, respectively, but inferior in terms of spatial scale ($1''.59$ pixel$^{-1}$ ) and temporal cadence ($\sim\,$5 min). 

The evolution of photospheric magnetic field in the source region is observed by the Helioseismic and Magnetic Imager \citep[HMI;][]{Scherrer2012} onboard SDO. In this study we worked with the Space-Weather HMI Active Region Patches (SHARP) vector magnetograms, which are de-projected to the heliographic coordinates with a Lambert (cylindrical equal area) projection method.

\subsection{Pre-Eruptive Evolution} \label{subsec:pre-eruption}

\begin{figure}[ht!]
\plotone{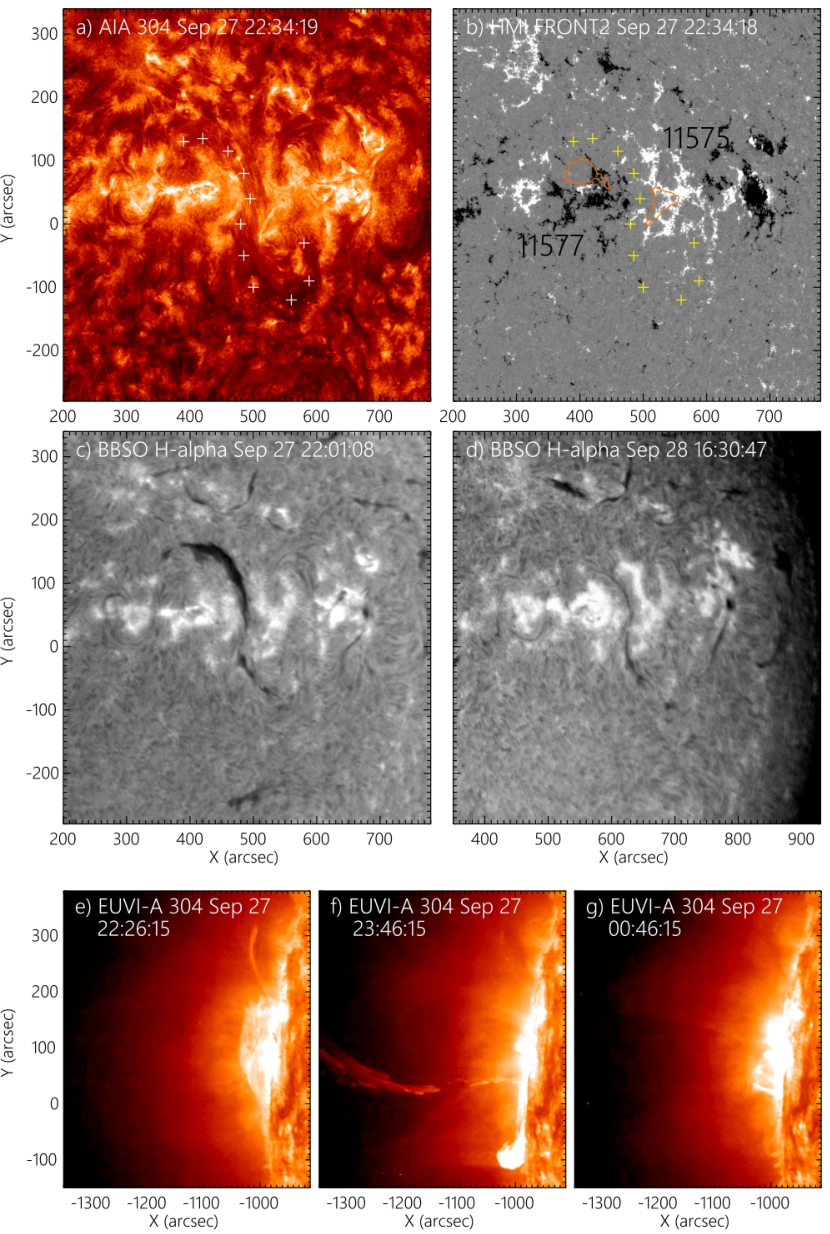}
\caption{Overview of the filament eruption on 2012 September 27. Panel (a) shows the sigmoidal filament observed in SDO/AIA 304~{\AA}. The corresponding line-of-sight HMI magnetogram is shown in (b), with active-region Nos. being labeled. The crosses mark the position of the filament channel. The two closed curves in orange indicate the conjugated dimming regions observed in AIA 304~{\AA} at about 23:50 UT (cf. Figure~\ref{fig:fps}a). Panels (c) \& (d) show the filament before and after the eruption in H$\alpha$ images taken by BBSO. (e--g) show the prominence before, during and after the eruption in 304~{\AA} images taken by STA/EUVI. Available online is an animation of AIA 304 and 131~{\AA} images with similar field of view as Panels (a--c), covering the time interval from 22:30 UT on 2012 September 27 to 00:20 UT on 2012 September 28.
\label{fig:overview}}  
\end{figure}

\begin{figure}[ht!]
	\plotone{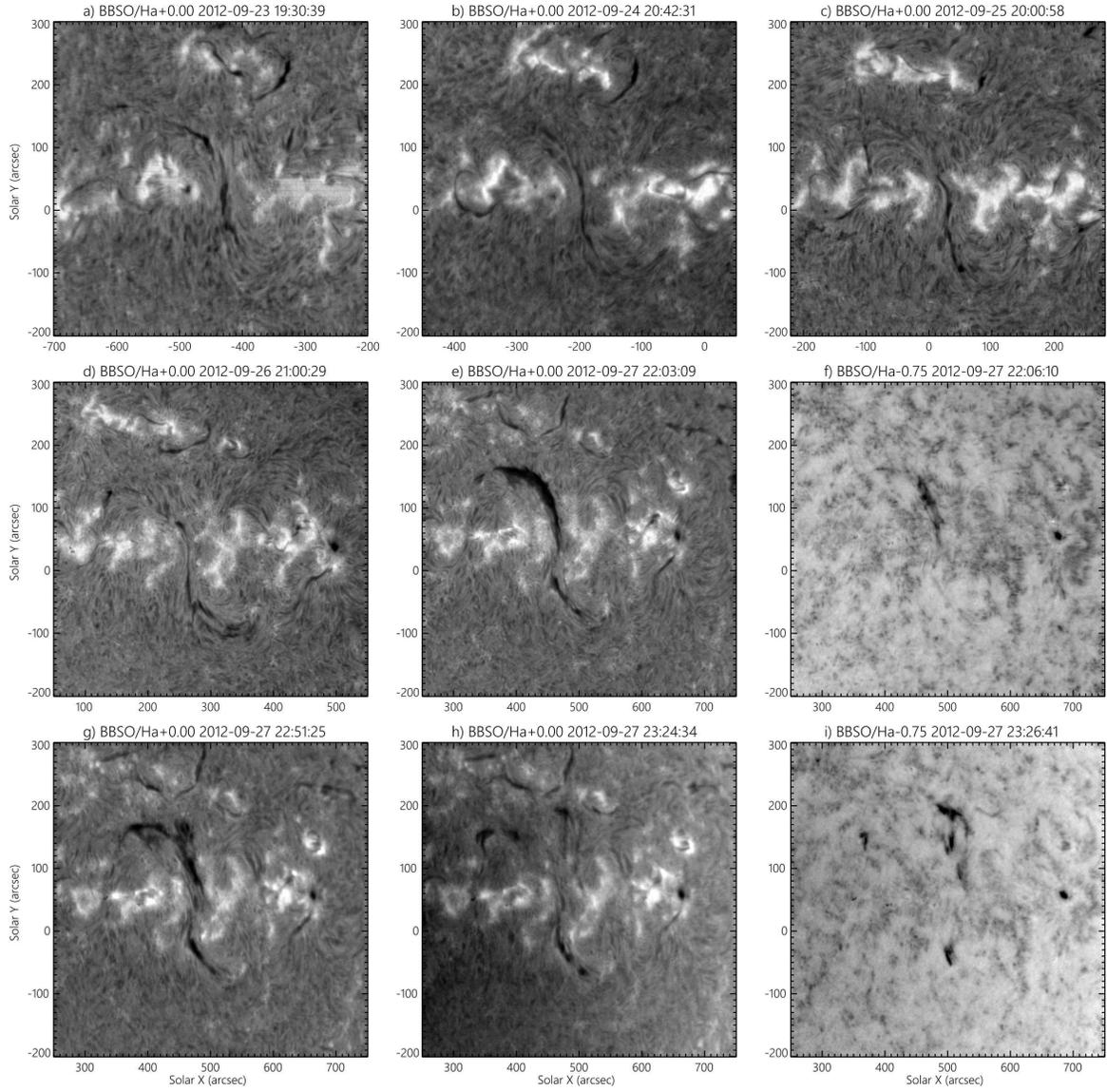}
	\caption{Snapshots of the filament in H$\alpha$ during the period from 2012 September 23 till 27. The images are taken in full disk by the Big Bear Solar Observatory, most in the H$\alpha$ line center, except two in blue wing (f \& i).    \label{fig:ha}}
\end{figure}

The filament of interest is an intermediate filament forming between two active regions. It has a reverse S shape (Fig.~\ref{fig:overview}a) aligned along the PIL that separates the trailing positive polarity of NOAA active region (AR) 11575 on the west from the leading negative polarity of AR 11577 on the east (Fig.~\ref{fig:overview}b). From ground-based H$\alpha$-filtergram observations one can see that the pre-eruption filament in H$\alpha$ (Fig.~\ref{fig:overview}c), especially its northern section, looks much thicker and darker than the post-eruption remnant (Figure~\ref{fig:overview}d), which is typical of partial eruptions. In fact, compared with its appearance during the earlier days, the filament looks darkest and thickest on 2012 September 27, just before its eruption (Figure~\ref{fig:ha}(a--e)).

From over one hour before the flare onset, the filament is observed to gradually split into two branches. Slowly rising upward at a projected speed of $\sim\,$5~\kms, the upper branch exhibits swirling motions about its long axis that has a north-south orientation, along with counterstreaming motions along the axis  (Figure~\ref{fig:precursor}(f \& g) and accompanying animation). Similar in 6 different AIA passbands including 131, 171, 193, 211, 304, and 335~{\AA}, the swirling appears clockwise if one looks along the field direction of the axis, i.e., from its southern end associated with positive polarity to northern end with negative polarity (cf. Figure~\ref{fig:overview}b). The swirling also gives an impression that the motion follows a right-handed screw and that the upper branch has a tubular shape whose northern section is broader.

The splitting causes weak brightenings at the splitting interface (see the inset in Figure~\ref{fig:precursor}d and the animation accompanying Figure~\ref{fig:overview}), which can be seen in both 304 and 131~{\AA}, suggesting that the filament material, originally dark in 304~{\AA}, is heated to transition-region temperatures. These brightened materials are transported by swirling motions back and forth, apparently along the tubular surface of the upper branch, which is manifested as the alternating black-and-white strips during 22:45-- 23:15 UT in a stack plot (Figure~\ref{fig:timeline}d and accompanying animation). The stack plot is made by taking a slice along a line segment (marked as `S1' in Figure~\ref{fig:precursor}d) from each 304~{\AA} image available during the specified time interval and then arranging the slices in chronological order. S1 has a length of 90~Mm and each slice is averaged over the slit width of 6~pixels. Such slits are hereafter referred to as  ``virtual slits'' to be differentiated from spectrograph slits.

From STA's viewing angle, one can only see the southern hooked part of the lower branch that is lying low near the surface; the rest of the lower branch is probably behind the limb. The upper branch, on the other hand, has risen high above the limb by 23:06 UT (Figure~\ref{fig:precursor}a). Its northern section is bent upward, while its middle section lies flat (Figure~\ref{fig:precursor}(a--b)). 
The northern section of the sigmoidal filament is activated as early as of 22:06 UT on September 27, as captured by an H$\alpha$ blue-wing image (Figure~\ref{fig:ha}f). At 22:51 UT, the splitting of the filament is visible in the H$\alpha$ center, but the upper branch appears to be still adhered to the lower branch (Figure~\ref{fig:ha}g). By the time of 23:26 UT, which is 10 min before the flare onset, the upper branch has completely detached from the lower branch (Figure~\ref{fig:ha}(h \& i)).

\subsection{Eruption Precursors} \label{subsec:precursor}
From about 23:25 UT onward, the swirling motions are replaced by more irregular and drastic motions, while the upper branch accelerates its peeling off from the lower branch and moves quickly westward in projection (see the animation accompanying Figure~\ref{fig:precursor}). As seen through the virtual slit S1, this sudden change of dynamics is also demonstrated as the disappearance of alternating black-and-white strips and an increase in speed from about 6~\kms to 24~\kms and a further increase to 68~\kms at the flare onset (Figure~\ref{fig:timeline}c). A J-shaped loop bundle is observed to rise synchronously with the upper branch and to disappear at the onset of the eruption (see the animation accompanying Figure~\ref{fig:overview}). The loop bundle outlines the northern section of the sigmoid; its northern footpoint is marked by a diamond in Figure~\ref{fig:precursor}d. The loop's rising speed increases from about 4~\kms to about 18~\kms at 23:25 UT in the stack plot (Figure~\ref{fig:timeline}e) made from a virtual slit S2 (Figure~\ref{fig:precursor}d), showing a similar evolution profile to the rising filament branch. Hence, it is considered to be part of the upper-branch structure.
	
Meanwhile, the upper branch of the filament is `whipped' southward from STA's perspective (Figure~\ref{fig:precursor}(b \& c)). While its southern leg is still attached to the surface, its northern leg has become too thin to be visible. From SDO's perspective, the ejection is not associated with any clear rotation of the filament axis (see the animation accompanying Figure~\ref{fig:precursor}). We measured the projected height at the horizontal section of the filament along a radial direction (Figure~\ref{fig:precursor}c). Despite the coarse cadence of EUVI images, the height-time plot shows a three-phase evolution in terms of the rising speed (Figure~\ref{fig:timeline}a), similar to what is observed by SDO. Associated with this southward whipping motion of the filament \citep[see also][]{Liu2009}, an arc-shaped wavefront appears at about 23:30 UT at the periphery of the source region, propagating southwestward, as observed in the difference images of the 193~{\AA} passband (Figure~\ref{fig:precursor}h). The wavefront is weak and diffuse, and disappears in a few minutes. 

\subsection{Eruptive Dynamics} \label{subsec:eruption}
As the flare onsets at 23:35 UT, two flare ribbons appear in the chromosphere, as observed in the AIA 304~{\AA} passband (see the animation accompanying Figure~\ref{fig:overview}). The two ribbons reside at each side of the PIL outlined by the filament (Figure~\ref{fig:overview}b), typical of the classic two-ribbon flares. With the progress of the flare, the two ribbons become thickened and separate from each other (Figure~\ref{fig:timeline}f), and the western (eastern) ribbon develops a `hook' at its southern (northern) end, enclosing a pair of coronal dimmings (labeled `DN' and `DS' in Figure~\ref{fig:precursor}e and Figure~\ref{fig:fps}a). The two J-shaped ribbons together constitute a reverse S shape, following a similar shape of the pre-eruption filament. The flaring loops that connect the two J-shaped ribbons are highly sheared in the south but potential-like in the north, therefore breaking the central symmetry of the reverse S-shape (Figure~\ref{fig:precursor}e). Specifically, in the wake of the eruption, the southern `elbow' of the S shape remains prominent, outlined by disturbed filament material draining towards the surface, which maps the southern footpoint of the remaining filament (labeled `LB$_\mathrm{sfp}$' in Figure~\ref{fig:fps}a); the northern `elbow' of the S shape becomes less recognizable with the presence of potential-like post-flare loops. This is because the eruptive branch mainly resides in the north, which is also evidenced by its footpoints manifested as the conjugated pair of coronal dimming. It is noteworthy that immediately following the draining at LB$_\mathrm{sfp}$ at about 23:55 UT, one can see a surge of dark, diffuse material in 304~{\AA}, which subsequently falls back toward the southern elbow (see the animation accompanying Figure~\ref{fig:overview}). Such a back bouncing of falling filament material is well observed in the renowned 2011 June 7 event and reproduced in two-dimensional hydrodynamic simulations by \citet{Reale2013}, assuming that magnetic field does not play an important role in such impacts on the surface. 

Enclosed by the southern hook, the dimming region DS can be clearly seen in multiple EUV passbands, including 304~{\AA}, 193~{\AA}, 171~{\AA}, 211~{\AA}, and 94~{\AA} (Figure~\ref{fig:fps}), a signature of mass loss associated with the eruption, which has been further corroborated through EUV spectroscopic analysis \citep{Veronig2019}. Enclosed by the northern hook, the dimming region DN is clearly visible in AIA 304~{\AA}, but obscure in other passbands. DS lies in the facular region of AR 11575, associated with relatively strong field, but DN lies mostly in the weak-field region to the north of AR 11577 \citep[see Figure~\ref{fig:overview}b and also][]{Veronig2019}. DS and DN together constitute a pair of conjugated footpoints of the eruptive structure.  A similar situation is found not only in the case of a forming flux rope, whose footpoints are identified as a pair of coronal dimmings enclosed by hooked flare ribbons \citep{Wang2017}, but also for the footpoints of four typical eruptive sigmoids \citep{Cheng&Ding2016}. It remains to be investigated what causes such an asymmetry in field strength in the footpoints of eruptive structures. It is noteworthy that DS begins to develop at about 23:30 UT, 5 min before the onset of the flare. This is demonstrated by the temporal variation of the average brightness (Figure~\ref{fig:timeline}a) sampled in a small box that is moving with the shifting and expanding dimming region (Figure~\ref{fig:fps}c). 

Although most of the source region observed in SDO/AIA is behind the solar limb from STA's viewing angle (Figure~\ref{fig:precursor}e), one can see in STA/EUVI 304~{\AA} that where the eruptive prominence's southern leg is connected to the surface is at the same latitude as DS (Figure~\ref{fig:precursor}(b \& e)), which provides a direct verification that DS is the southern footpoint of the erupting structure. Recall that the J-shaped loop bundle rises synchronously with the upper branch (\S\ref{subsec:precursor}). Its northern footpoint (labeled `J-loop$_\mathrm{nfp}$' in Figure~\ref{fig:fps}a) falls within the northern dimming region DN, which corroborates our conclusion that DN is the northern footpoint of the erupting structure.

\subsection{Evolution of the Source Region}
\begin{figure}[ht!]
	\plotone{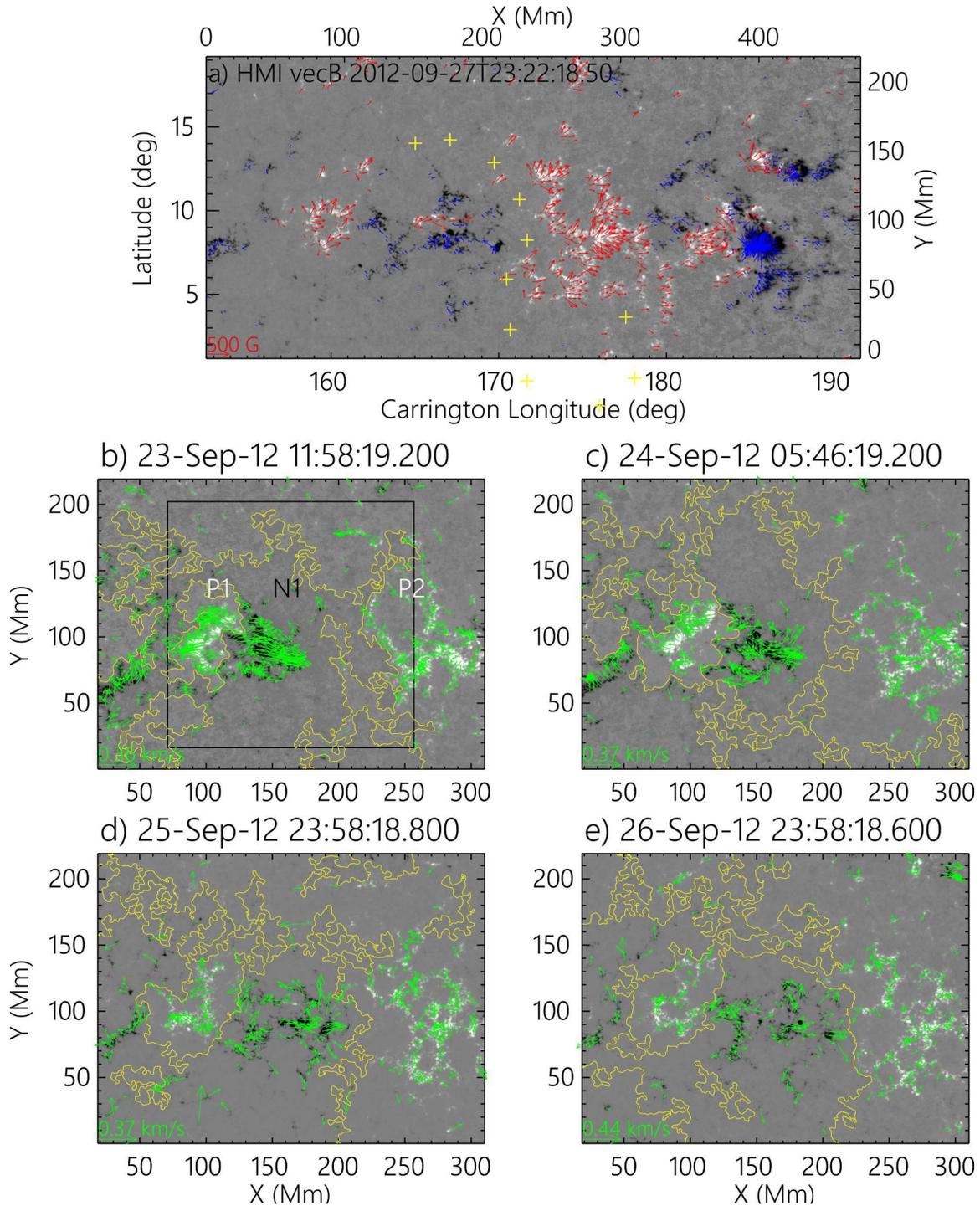}
	\caption{HMI observations of AR 11577 and 11575. The filament of interest is located in between the two active regions. (a) Vector magnetogram taken at 23:22 UT on 2012 September 27, immediately before the flare onset at 23:36. White (black) indicates positive (negative) $B_z$, which is saturated at $\pm800$ G. Red (blue) arrows represent the tangential field components that originate from positive (negative) $B_z$. The yellow crosses that mark the position of the filament channel in Figure~\ref{fig:overview}b are converted to Carrington longitude and latitude and replotted here. (b--d) Maps of photospheric flows superimposed on maps of $B_z$. Each flow map is averaged over 2 hrs centering on the time of the $B_z$ map. In the lower left corner, the horizontal arrow indicates the maximum of flow vectors shown in the map. The rectangle in (b) marks the region in which the helicity injection is derived (Figure~\ref{fig:ltc}). The major PILs (yellow) are indicated by the contours of $B_z=0$. Contours with relatively short lengths have been discarded. \label{fig:bfield}} 
\end{figure}

\begin{figure}[ht!]
	\plotone{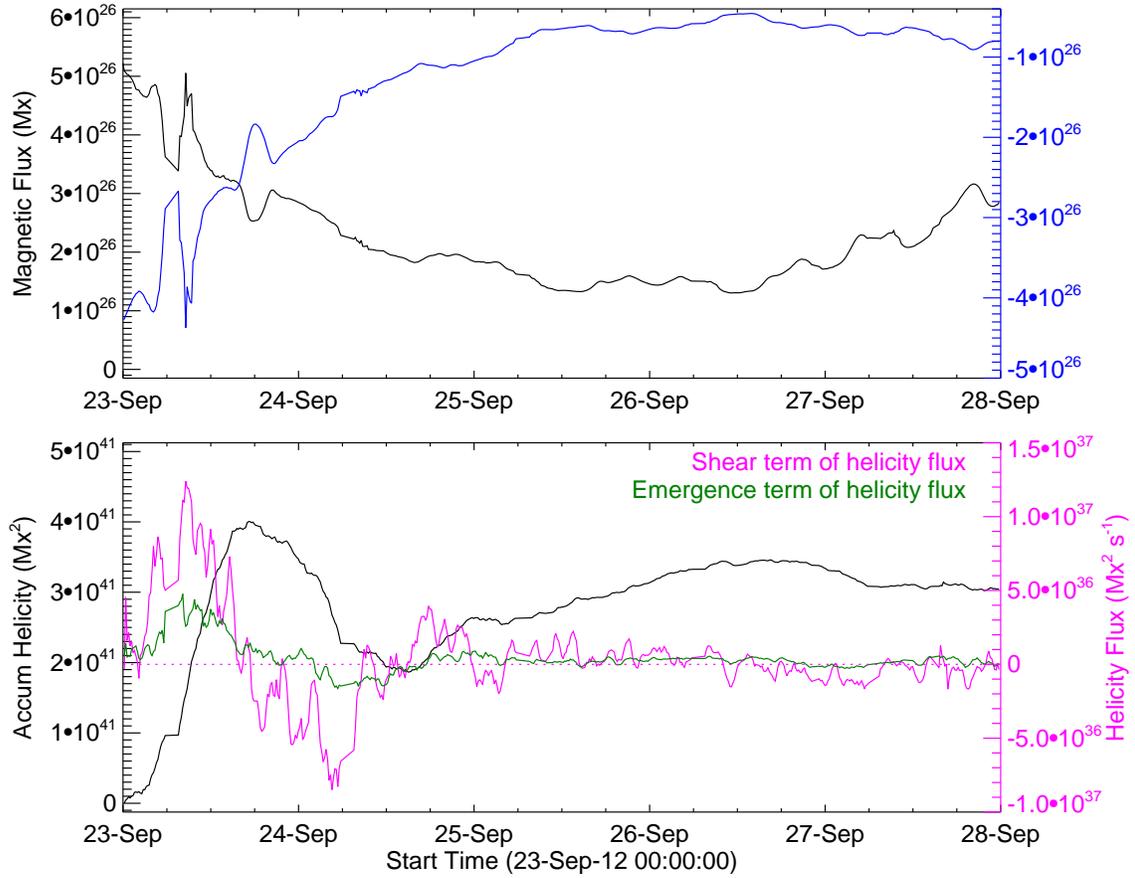}
	\caption{Temporal evolution of magnetic flux and helicity flux through the region of interest. Top: magnetic flux in units of Maxwell (Mx) through the same field of view as in Figure~\ref{fig:bfield}(b--e). Positive (negative) flux is shown in black (blue) and scaled by the left (right) $y$-axis. Bottom: Accumulative helicity in units of Mx$^2$, which is injected into the atmosphere through the rectangle marked in Figure~\ref{fig:bfield}b. The helicity flux in units of Mx$^2$~s$^{-1}$ is scaled by the right $y$-axis, with magenta and green indicating the contribution from photosheric motions and flux emergence, respectively. The curves in this figure are smoothed by a running box of five data points. \label{fig:ltc}}
\end{figure}

At this point we look back at the evolution of the source region leading up to the filament eruption. The filament is formed between AR 11577 and 11575. Both are decayed and categorized as $\beta\gamma$ according to the Mount Wilson classification as of the time of eruption. However, AR 11577 acquires its active-region number starting only from 2012 September 23, as an emerging active region with two major sunspots N1 and P1 moving westward and eastward, respectively (Figure~\ref{fig:bfield}b). Meanwhile, the flux concentrations become scattered, and in the field of view specified by Figure~\ref{fig:bfield}(b--e) both negative and positive flux decrease in magnitude with time (top of Figure~\ref{fig:ltc}). By late September 25 (Figure~\ref{fig:bfield}d), N1 has approached P2, the positive-polarity elements from AR 11575. 

With the time-series of deprojected, co-registered HMI vector magnetograms (e.g., Figure~\ref{fig:bfield}a), we employed the Differential Affine Velocity Estimator for Vector Magnetograms \citep[DAVE4VM;][]{Schuck2008} to obtain the flow field (e.g., Figure~\ref{fig:bfield}(b--e)) and further calculated the relative helicity flux across the photospheric boundary $S$ as follows \citep{Berger1984},
\begin{equation}
\left.\frac{dH}{dt}\right|_S=2\int_S(\mathbf{A}_p\cdot\mathbf{B}_t)V_{\perp n}\,dS-2\int_S(\mathbf{A}_p\cdot\mathbf{V}_{\perp t})B_n\,dS, \label{eq:helicity}
\end{equation}
in which $t$ and $n$ indicate the tangential and normal directions, respectively. $\mathbf{A}_p$ is the vector potential of the reference potential field that is based on the photospheric $B_n$. $\mathbf{V}_\perp = \mathbf{V} - (\mathbf{V}\cdot\mathbf{B})\mathbf{B}/B^2$ is the photospheric velocity perpendicular to magnetic field lines \citep{Liu&Schuck2012}. The helicity injection into the active region can be attributed to either flux emergence (the term with $V_{\perp n}$) or photospheric motions that shear and braid field lines (the term with $V_{\perp t}$). 

Within a rectangular region that includes N1, P1, and part of P2 (Figure~\ref{fig:bfield}b), the shear term with $\mathbf{V}_{\perp t}$ dominates over the emergence term with $V_{\perp n}$ until early September 24 (bottom of Figure~\ref{fig:ltc}). An episode of positive helicity injection on September 23 is associated with the emergence of the paired sunspots N1 and P1 in AR 11577, which is followed by an episode of negative helicity injection on September 24, associated with the migration of N1 toward P2 in AR 11575 (Figure~\ref{fig:bfield}). After that, contributions to helicity injection from both flux emergence and photospheric motions become negligible. Eventually, although the accumulated helicity is positive by September 27, its magnitude ($\sim10^{41}$~Mx$^2$) is one order below the typical helicity accumulated in emerging active regions \citep[e,g,][]{LiuY2014,Liu2016}. The filament of interest is already present to the east of AR 11575 before the emergence of AR 11577 (not shown), and maintains an inverse S shape until its eruption on September 27 (Figure~\ref{fig:ha}), indicating that the chirality (helicity) of the filament is not significantly affected by the evolution of AR 11577.  

What dominates the evolution of the source region from September 25 onward is the flux cancellation between N1 and P2, the dispersal of P1, and the converging motion of magnetic elements toward the PIL between N1 and P2 at speeds below 0.5~\kms from either side of the PIL (Figure~\ref{fig:bfield}(d \& e)). It is well known that a flux rope can be formed through tether-cutting-like reconnections in a magnetically sheared arcade. The reconnection can be driven either by dispersal and diffusion of photospheric fluxes or by shearing and converging flows around the PIL \citep[e.g.,][]{vanBallegooijen&Martens1989,Moore2001,Green2011,Aulanier2010}. From September 23 to 26, the flux cancellation rate is roughly $10^{26}$~Mx per day (top panel of Figure~\ref{fig:ltc}). But on September 26, both positive and negative magnetic flux become ``saturated'', changing little in magnitude with time. Starting from September 27 onward, magnetic flux even increases slightly in magnitude with time. Thus, with the diminishing rate of flux cancellation, we consider it likely that the magnetic structure associated with the sigmoidal filament has taken on a definite form by September 26.

\section{Discussion \& Conclusion} \label{sec:disc}
To summarize, we investigated the splitting of a sigmoidal double-decker filament, which experiences a three-stage evolution, i.e., a quasi-static stage, a precursor stage, and the partial eruption. During the quasi-static stage, the filament splits into two branches starting from over 1.5 hr before the eruption. The splitting proceeds gradually, during which the upper branch rises at a few kilometers per second. Meanwhile, the upper branch exhibits swirling motions, with weak heating at the splitting interface. During the precursor stage, which lasts about 10 min before the eruption, GOES 1--8~{\AA} flux is slightly enhanced (Figure~\ref{fig:timeline}a), the rising speed of the upper branch increases to tens of kilometers per second from both SDO and STB perspectives, and the regular swirling motions within the upper branch are replaced by somewhat turbulent motions, while a pair of conjugated dimming regions start to develop at the footpoints of the upper branch. Meanwhile, the southward whipping motion of the upper branch drives an arc-shaped EUV front propagating in the similar direction. Finally during the eruption stage as characterized by the separation of two flare ribbons, the upper branch is expelled with a rapid increase in speed to hundreds of kilometers per second, while the lower branch maintaining a sigmoidal shape is left behind.

\subsection{Pre-eruption Structure} \label{subsec:Pre-eruptionStruc}
This observation demonstrates that a double-decker filament may form ``quiescently'' through gradual photospheric evolution, but only split to display two well-separated branches immediately before the eruption. In our case the splitting process begins as early as 1.5 hr before the onset of the flare. The two branches of the double-decker filament must be embedded in two distinct flux bundles before the eruption. This is evidenced by the distinctive separation between the southern footpoint of the remaining filament and that of the eruptive structure (Figure~\ref{fig:fps}a). 

The footpoints of the eruptive structure are identified through the conjugated dimming regions DN and DS, the latter of which starts to develop due to the enhancement in rising speed during the precursor stage (Figure~\ref{fig:fps}c). The remaining filament keeps roughly the same sigmoidal shape in H$\alpha$ (Figure~\ref{fig:overview}d), its northern footpoint must be close to the northern end of the filament channel (Figure~\ref{fig:overview}a; labeled `FL$_\mathrm{ne}$' in Figure~\ref{fig:fps}a), just like its sourthern footpoint, as highlighted by draining filament material, is close to the southern end of the filament channel (labeled `FL$_\mathrm{se}$' in Figure~\ref{fig:fps}a). Since this separation in footpoints between the eruptive and remaining branch does not result from magnetic reconnection during the eruption but is rather `pre-eruptive', the double-decker structure must be already present in the apparently single filament long before the eruption. The two flux bundles might be adhered to each other before splitting, similar to the simulation in \citet{Kliem2014}. The heating of the splitting interface to transition-region temperatures might be related to the current accumulation and dissipation at the HFT separating the two flux bundles. 

The swirling motion and the tubular shape as the motion reveals argue strongly for a flux-rope configuration possessed by the upper branch. However, it is unclear whether the swirling reflects mass motions along twisted field or motions of twisted field itself. On the other hand, reverse S-shaped structures are normally associated with negative helicity \citep{Green2007,Zhou2020}. Further, footpoints of both the upper and lower branch are clearly left-skewed relative to the PIL, which is also a signature of negative helicity \citep{Chen2014,Ouyang2017}. 

Thus, if the rope is left-handed, the clockwise swirling (see \S~\ref{subsec:pre-eruption}) may result from the untwisting of helical field lines, which can be naturally attributed to the decrease in twist density in a flux rope under rise and expansion.  It is also possible that part of the twist is transferred to the ambient flux through reconnection between the flux rope and the ambient field, a mechanism proposed for the untwisting motions observed occasionally during the main eruption phase \citep[e.g.,][]{Xue2016}. Here due to the absence of obvious reconnection signatures associated with the swirling motions, we surmise that the reconnection rate must be extremely small if this mechanism were indeed at work.	
	
\subsection{Eruption Mechanism}  \label{subsec:EruptionMech}
Judging from images taken from two different viewing angles, one can see that the eruptive process does not involve any clear kinking motions \cite[e.g.,][]{Gilbert2007}, which excludes the helical kink instability as the trigger of the eruption. 

\begin{figure}[ht!]
	\plotone{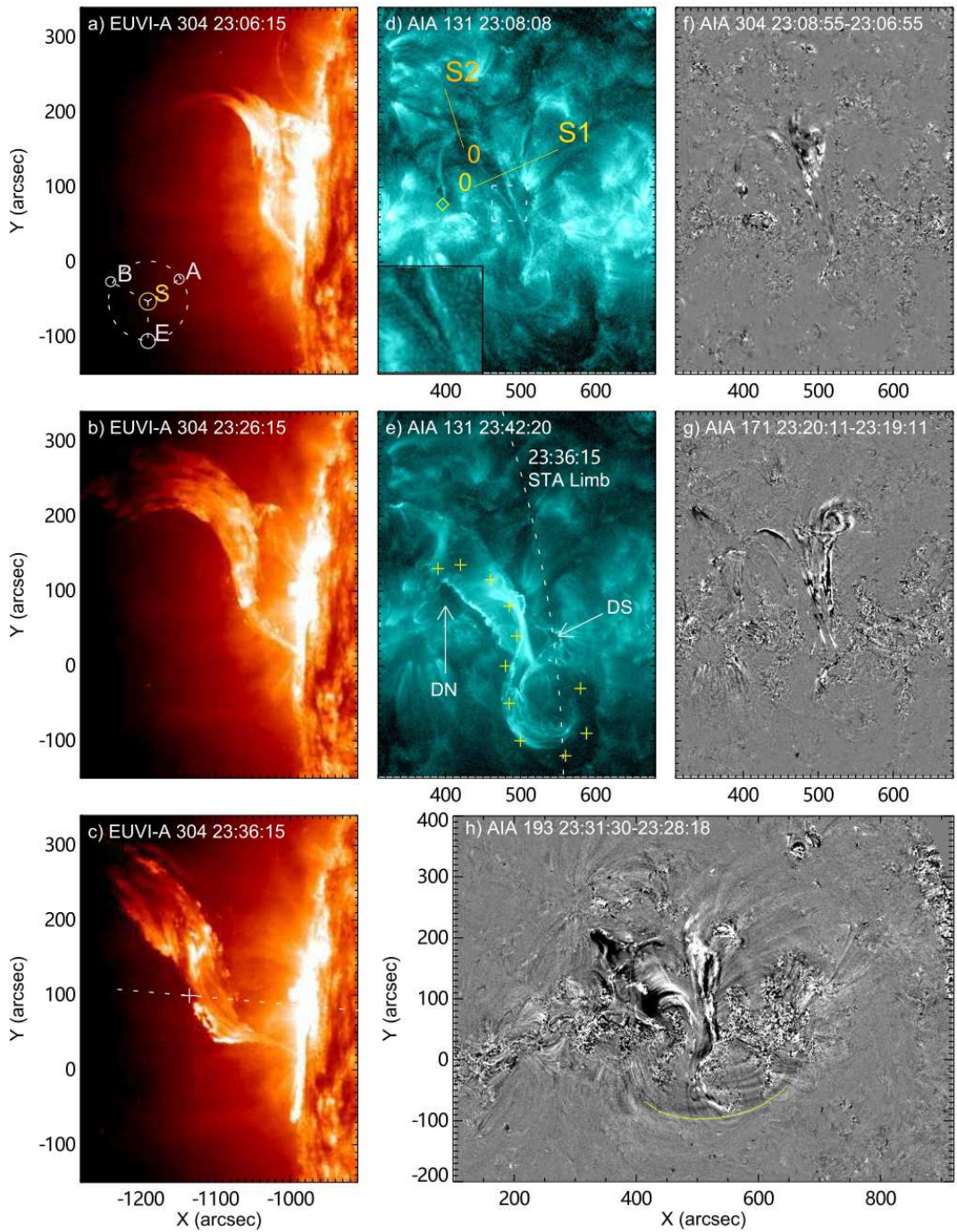}
	\caption{Precursor stage of the filament eruption. Panels (a--c) show the prominence above the limb as observed in 304~{\AA} by STA/EUVI. The inset in (a) indicates the positions of two STEREO satellites (A and B) relative to the Sun (S) and Earth (E) in the solar ecliptic plane. The dotted line in (c) indicates the direction along which the prominence height (marked by a cross) is measured. Panels (d \& e) show the corresponding filament on the disk as observed in 131~{\AA} by SDO/AIA. Two virtual slits S1 and S2 are marked in (d), with `0' indicating the where each slit starts. The inset in (d) enlarges a rectangular region at the center of the filament, where it bifurcates. The footpoint of a J-shaped loop bundle is marked by a diamond. Crosses in (e) mark the position of the filament channel as shown in Figure~\ref{fig:overview}a. The dotted curve indicates the solar limb from STA's perspective. DN and DS mark two dimming regions. Panels (f--h) show difference images in AIA 304, 171, and 193~{\AA}, respectively. The yellow arc in (h) outlines an EUV front propagating southwestward. Available online is an animation of running difference images from six AIA passbands, 131, 171, 193, 211, 304, and 335~{\AA}, with the same field of view as Panels (d--g), covering the time interval from 22:30 to 23:50 UT on 2012 September 27.  \label{fig:precursor}} 
\end{figure}

\begin{figure}[ht!]
	\epsscale{0.9}
	\plotone{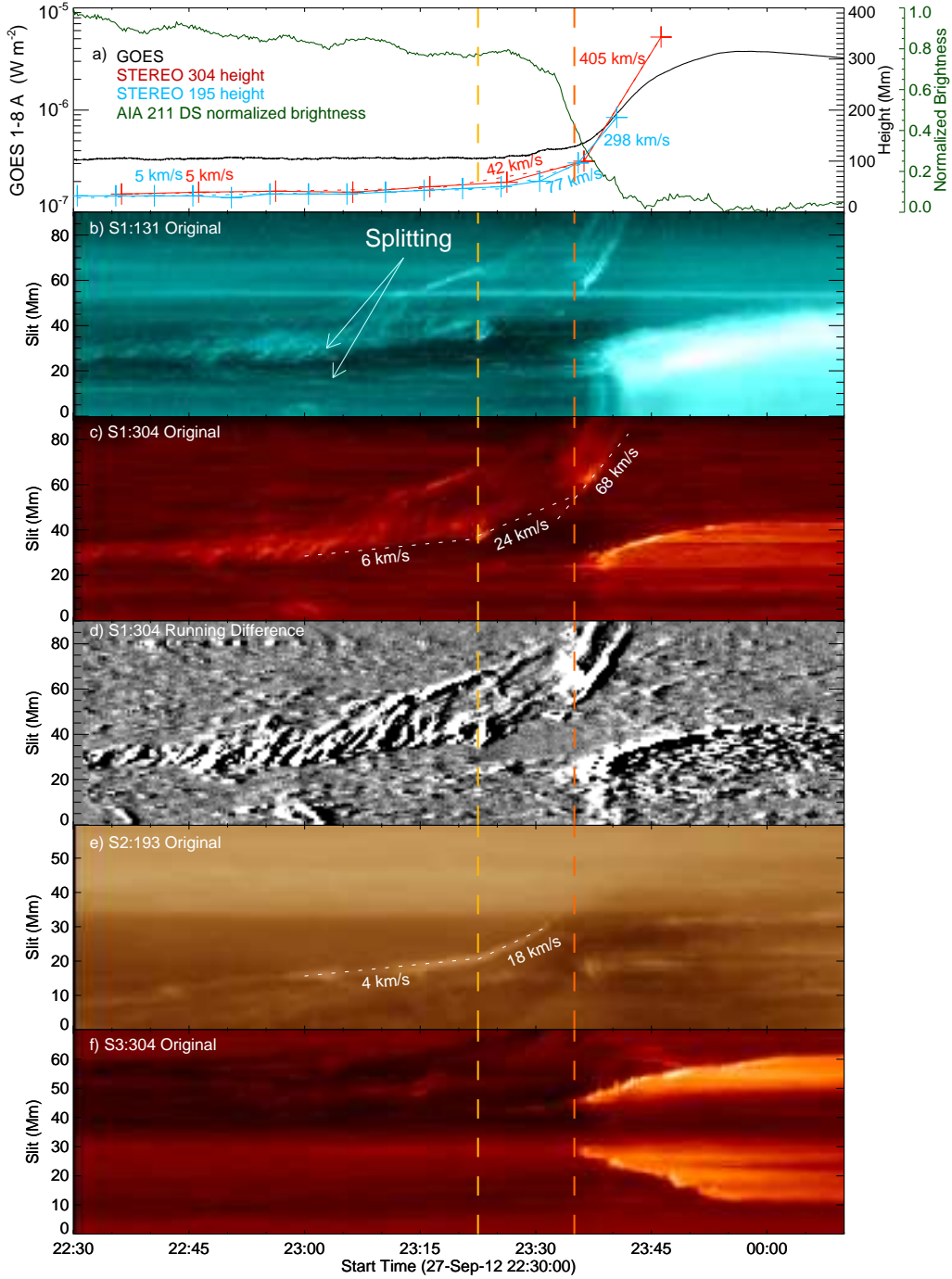}
	\caption{Kinematics of the double-decker filament in relation to the flare. Panel (a) shows the GOES 1--8~\AA\ flux as scaled by the left $y$-axis, projected heights of the prominence in STA/EUVI 304 (red) and 195~{\AA} (blue) as scaled by the right $y$-axis in black, and the normalized brightness in the south dimming DS (green) as scaled by the right $y$-axis in green. Linear fits of the height-time plots give an estimation in speeds at different time durations. Panels (b--d) show the evolution of the filament seen through the virtual slit S1 (marked in Figure~\ref{fig:precursor}d). The stack plots are made from SDO/AIA 131~{\AA}, 304~{\AA}, and the running difference of 304~{\AA} images, respectively. Panel (e) shows the evolution of a bundle of J-shaped loops in SDO/AIA 193~{\AA} seen through the virtual slit S2 (marked in Figure~\ref{fig:precursor}d). Panel (f) shows the evolution of flare ribbons in SDO/AIA 304~{\AA} seen through the virtual slit S3 (marked in Figure~\ref{fig:fps}a). The two vertical dashed lines indicate the beginning of the precursor stage and of the partial eruption, respectively. Available online is an animation of AIA 131~{\AA}, 304~{\AA}, and the running difference of 304~{\AA} images; the slice (S1) taken from each image is indicated on the associated stack plots (corresponding to panels (b--d), respectively).  \label{fig:timeline}} 
\end{figure}

\begin{figure}[ht!]
\plotone{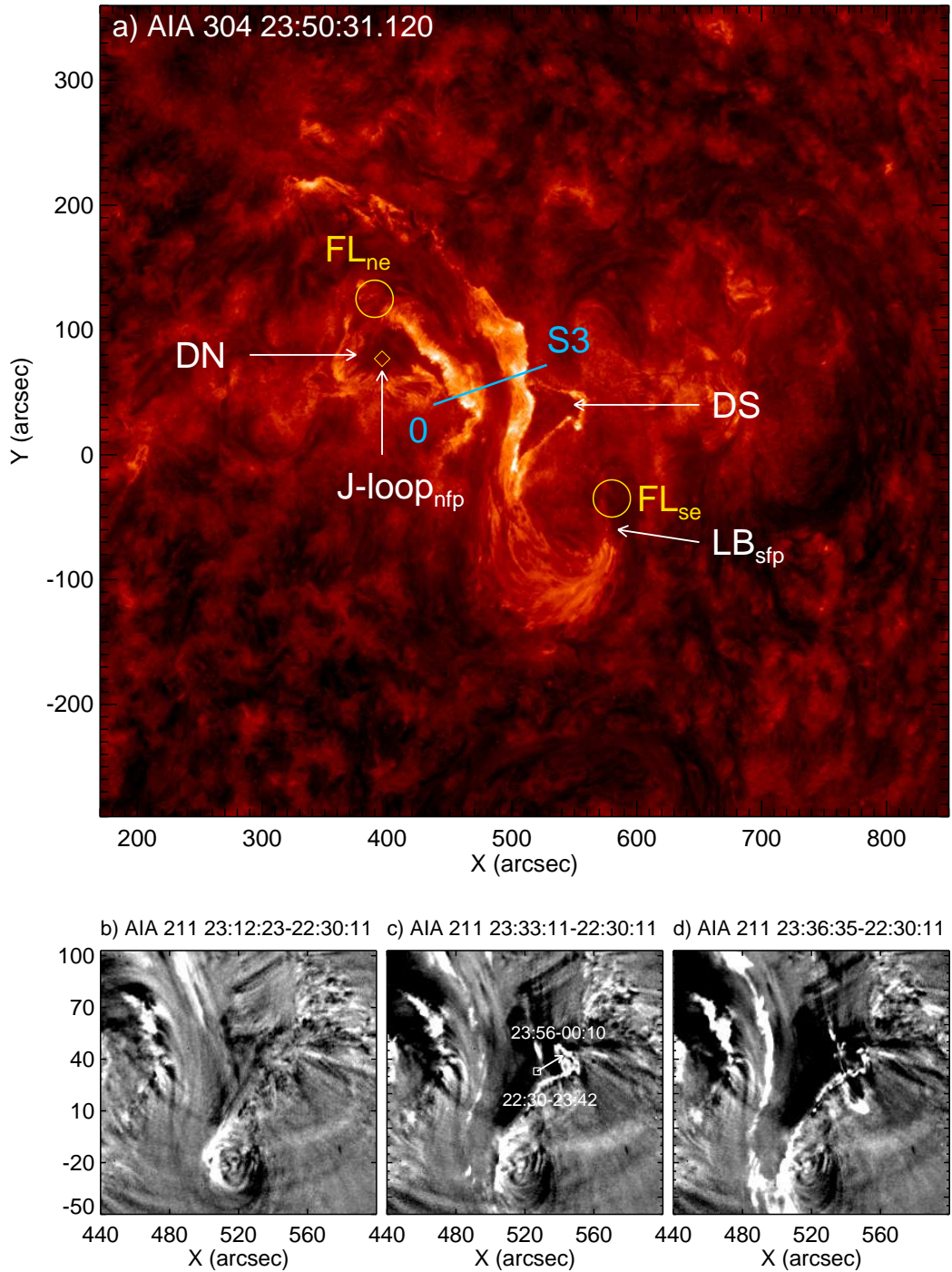}
\caption{Footpoints of the eruptive and remaining structure. In Panel (a), `DN' and `DS' indicate the conjugated pair of coronal dimmings; `LB$_\mathrm{sfp}$' indicates the southern footpoint of the remaining filament; the two circles labeled `FL$_\mathrm{ne}$' and `FL$_\mathrm{se}$' mark the northern and southern end of the filament channel as observed in 304~{\AA} before the eruption (cf. Figure~\ref{fig:overview}a); the diamond labeled `J-loop$_\mathrm{nfp}$' marks the northern footpoint of the J-shaped loop bundle as identified in Figure~\ref{fig:precursor}d. A virtual slit S3 is placed across two flare ribbons. The slit is of 70 Mm long, and the label '0' indicates where it starts. The resultant stack plot is shown in Figure~\ref{fig:timeline}f. Panels (b--d) zoom into the dimming region to show its evolution in 211~{\AA} running difference images. The temporal variation in average brightness of pixels in a small rectangle in (c) is shown in Figure~\ref{fig:timeline}a. The rectangle is made to move along a direction indicated by an arrow, as the dimming region shifts and expands northwestwards. \label{fig:fps} }
\end{figure}

The prominence's flat part has risen to as high as 80~Mm above the surface at the flare onset (Figure~\ref{fig:timeline}a), its arched part is twice as much high at the same time (Figure~\ref{fig:precursor}c). Following the procedure in \citet{WangD2017}, we found that the decay index $n=-d\ln B/d\ln h$ at $\sim\,$80~Mm above the photospheric PIL is $3.4\pm0.6$, and alternatively $n$ at the PIL of the potential field at $\sim\,$80~Mm is $2.3\pm0.2$; both are significantly above the theoretical threshold value of 1.5 \citep{Kliem&Torok2006}. Hence, the torus instability may be dominant in the eruption of the upper branch. 

In conclusion, the double-decker filament under investigation possesses two distinct flux bundles, as demonstrated not only by the two well-separated branches but also by their well-separated footpoints during the precursor and eruption stages. The formation of the double-decker structure is associated with gradual magnetic flux cancellations and converging photospheric flows around the PIL over a few days. The splitting of the two flux bundles before eruption lifts the upper branch to a torus-unstable regime and result in the partial filament eruption. Thus, alternative to the reconnection-driven splitting during eruptions, this scenario provide a new evolutionary path towards partial eruptions of solar filaments. Such filaments may appear single but be actually embedded in a double-decker structure already before eruption \citep[e.g., see also][]{Cheng2014,Awasthi2019}. The pre-eruption splitting is therefore not only an inherent process in the double-decker structure but also an indicator of the imminent eruption.

\acknowledgements  
This work was supported by the National Natural Science Foundation of China (NSFC; Grant Nos. 41761134088, 41774150, 11925302, and 11903032) and the Strategic Priority Program of the Chinese Academy of Sciences (Grant No. XDB41030100). BK acknowledge the NSFC-DFG collaborative grant KL 817/8-1.


\end{document}